\documentclass[amsmath, amsfonts, twocolumn, prl]{revtex4}
\usepackage{graphicx}
\usepackage{epsfig}
\usepackage{bm}
\usepackage{dcolumn}
\usepackage{amsmath}
\usepackage{amssymb}
\usepackage{color}
\usepackage[varg]{txfonts}

\newcommand{\dett}{\mathop{\rm Det}\nolimits}

\begin{document}

\title{Topological Andreev bands in three-terminal Josephson junctions}

\author{Hong-Yi Xie}
\affiliation{Department of Physics, University of Wisconsin-Madison, Madison, Wisconsin 53706, USA}

\author{Maxim G. Vavilov}
\affiliation{Department of Physics, University of Wisconsin-Madison, Madison, Wisconsin 53706, USA}

\author{Alex Levchenko}
\affiliation{Department of Physics, University of Wisconsin-Madison, Madison, Wisconsin 53706, USA}

\date{August 29, 2017}

\begin{abstract}
We study the emergent band topology of subgap Andreev bound states in the three-terminal Josephson junctions. We scrutinize the symmetry constraints of the scattering matrix in the normal region connecting superconducting leads that enable the topological nodal points in the spectrum of Andreev states. When the scattering matrix possesses time-reversal symmetry, the gap closing occurs at special stationary points that are topologically trivial as they carry vanishing Berry fluxes. In contrast, for the time-reversal broken case we find topological monopoles of the Berry curvature and corresponding phase transition between states with different Chern numbers. The latter is controlled by the structure of the scattering matrix that can be tuned by a magnetic flux piercing through the junction area in a three-terminal geometry. The topological regime of the system can be identified by nonlocal conductance quantization that we compute explicitly for a particular parametrization of the scattering matrix in the case where each reservoir is connected by a single channel. 
\end{abstract}

\maketitle

\textbf{\textit{Introduction}}. The Wigner-Dyson classes of Gaussian random-matrix ensembles of orthogonal, unitary, and symplectic symmetry \cite{Wigner,Dyson,Mehta} play a central role in mesoscopic physics, as they describe the universal ergodic limit of disordered and chaotic single-particle systems. The power of such random-matrix theory (RMT) description is that it enables predictive statements about the properties of a system, such as level statistics and level correlations, transport conductance and its fluctuations, etc., by circumventing the need for a microscopic description of the system \cite{Beenakker-RMP}. Study of normal-superconductor hybrid mesoscopic devices carried out by Altland and Zirnbauer \cite{Altland} led to the extension in applications of RMT phenomenology in solid-state systems to include nonstandard Cartan symmetry spaces. This work paved the way for a complete classification of gapped phases of noninteracting fermions \cite{Qi,Schnyder,Ryu,Kitaev} (see also recent reviews \cite{Ludwig,Chiu}). In any given spatial dimension only five of the ten symmetry classes host topologically nontrivial phases. The topology can be identified as a mapping from the properties of bands in the Brillouin zone to a certain integral invariant such as a Chern number \cite{Chern}. 

The early surge for band topology was concentrated around various lattice models: Haldane \cite{Haldane}, Kane-Mele \cite{Kane-Mele}, Bernevig-Hughes-Zhang \cite{BHZ}, and Kitaev \cite{Kitaev-Model}, which has since been expanded to include crystalline symmetries \cite{Zaanen}. These initial ideas, spread across different physics disciplines and topological properties, are being discovered and extensively studied beyond crystals. The list of examples includes photonic arrays, coupled resonators, metamaterials and quasicrystals \cite{Lu}, colloids \cite{Lubensky}, and even amorphous media \cite{Agarwala}, while some of the proposed lattice Hamiltonians were realized with cold atoms \cite{Esslinger}. 

\begin{figure}
\includegraphics[width=0.23\textwidth]{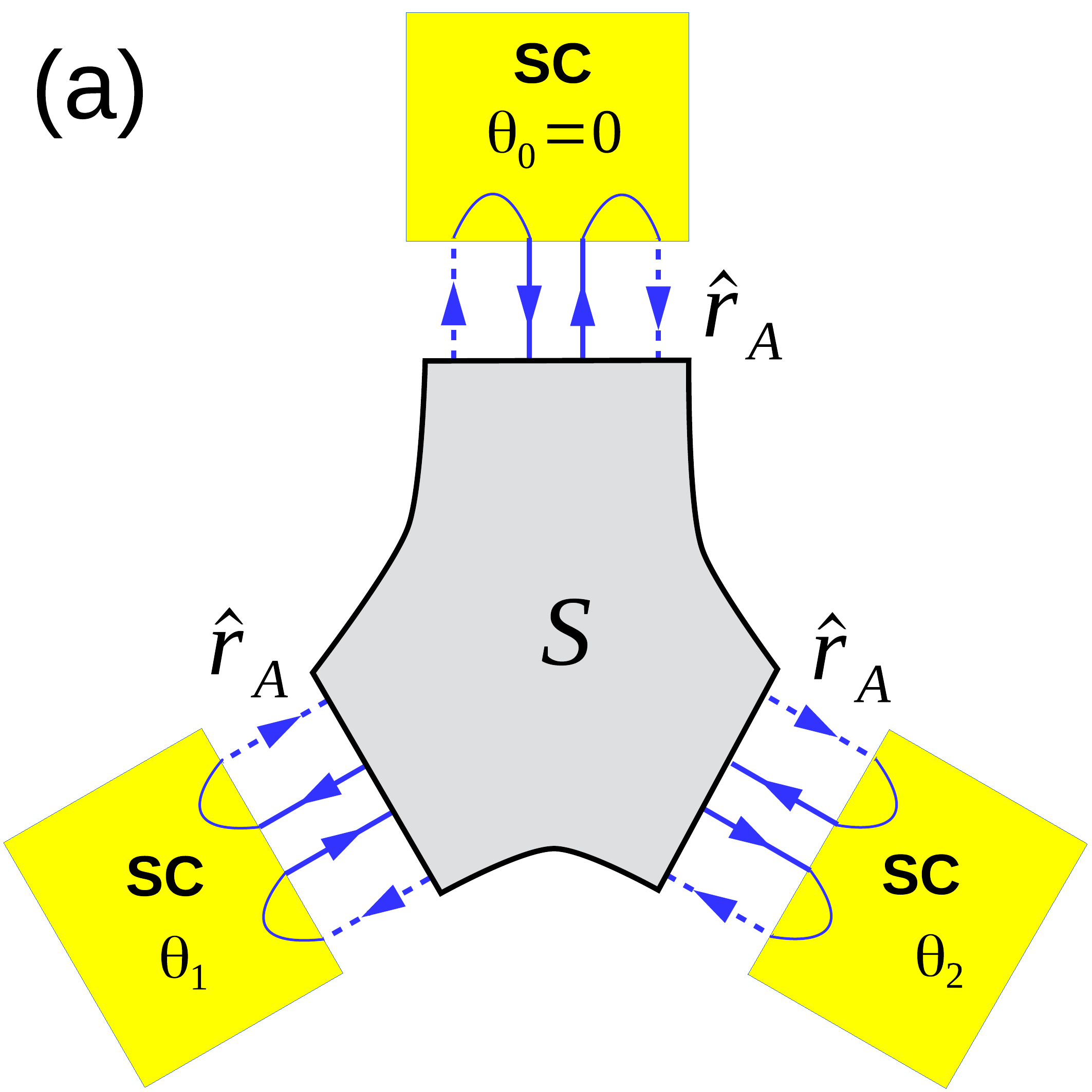}
\includegraphics[width=0.23\textwidth]{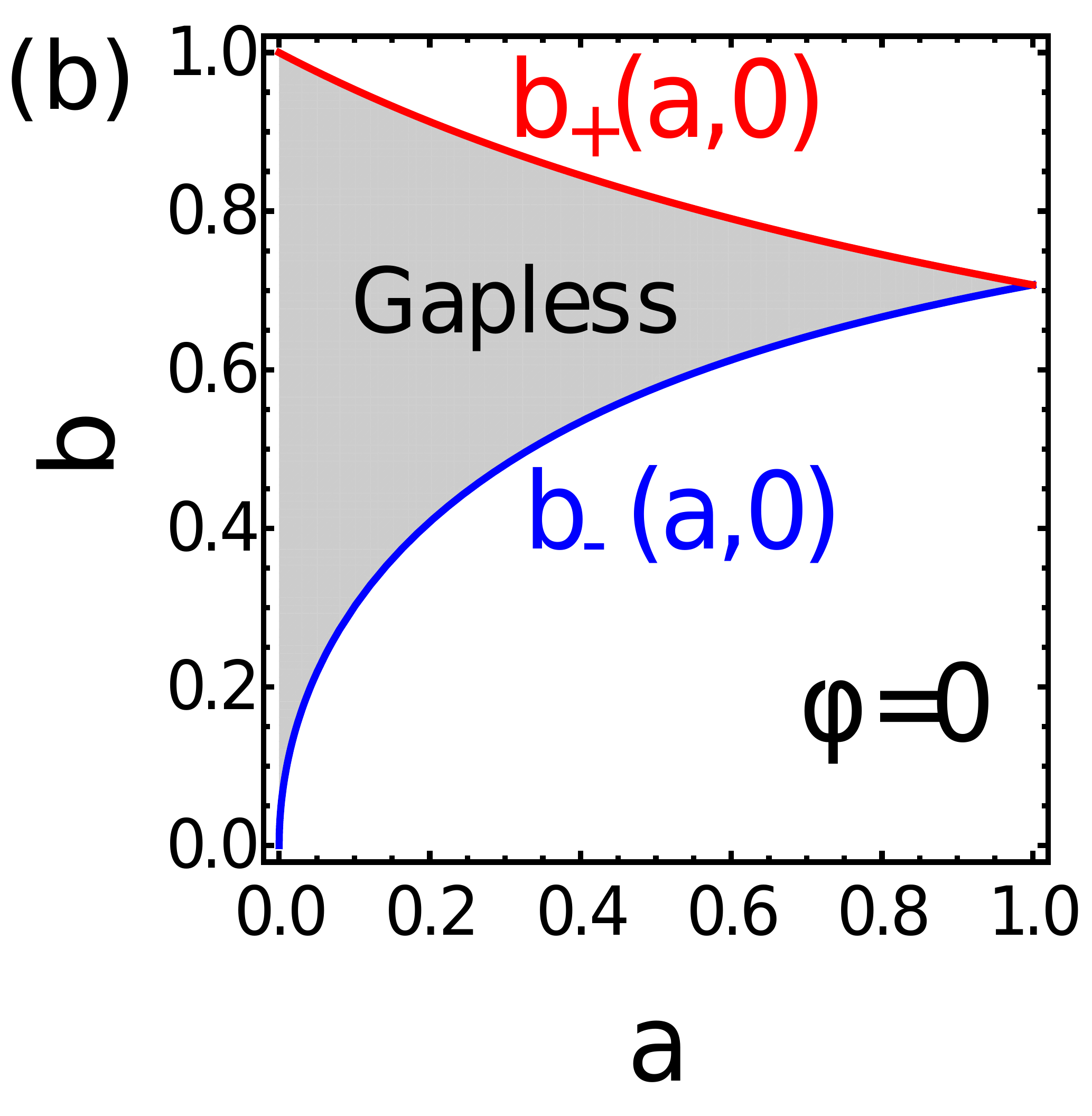} \\
\includegraphics[width=0.23\textwidth]{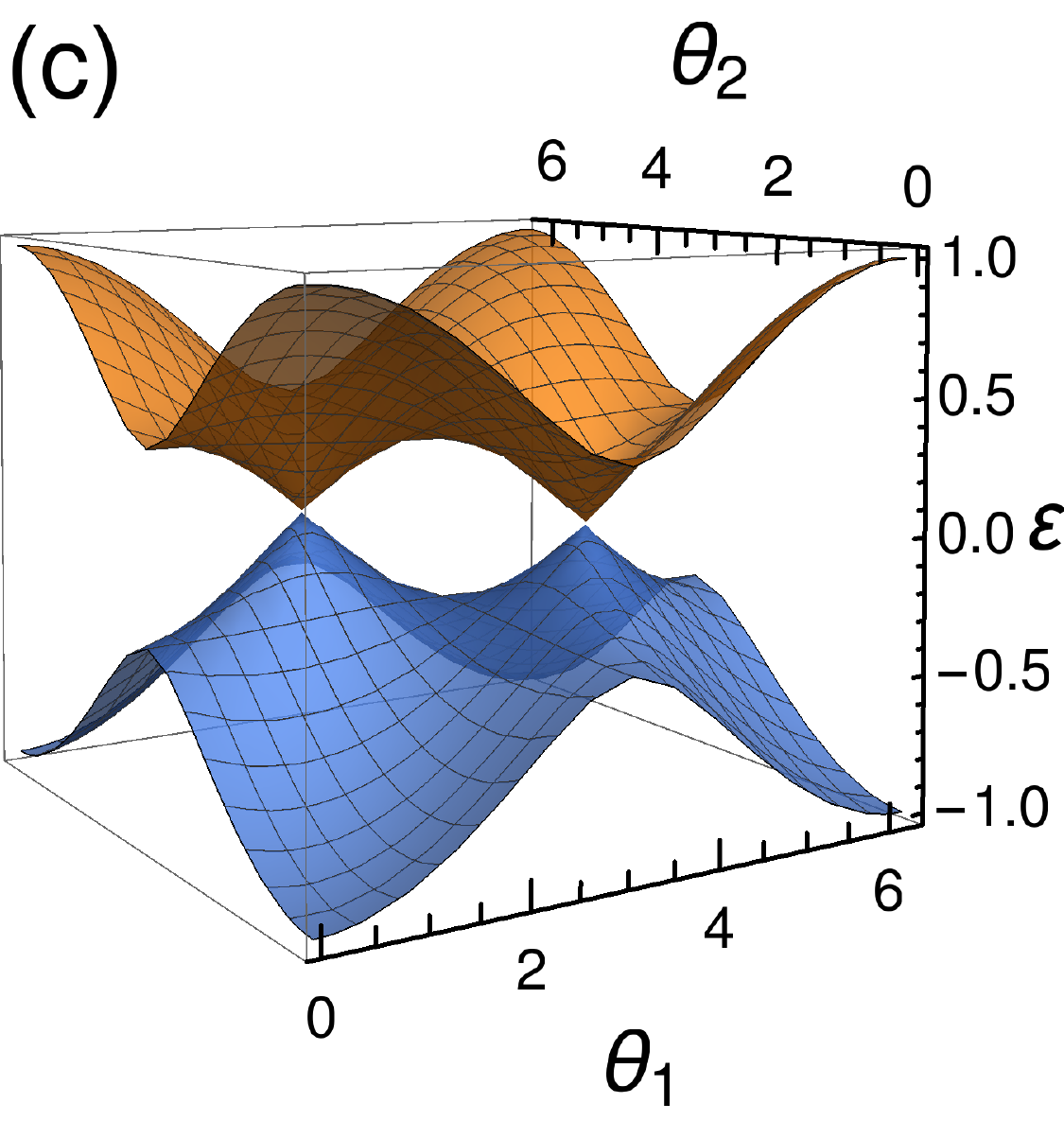}
\includegraphics[width=0.23\textwidth]{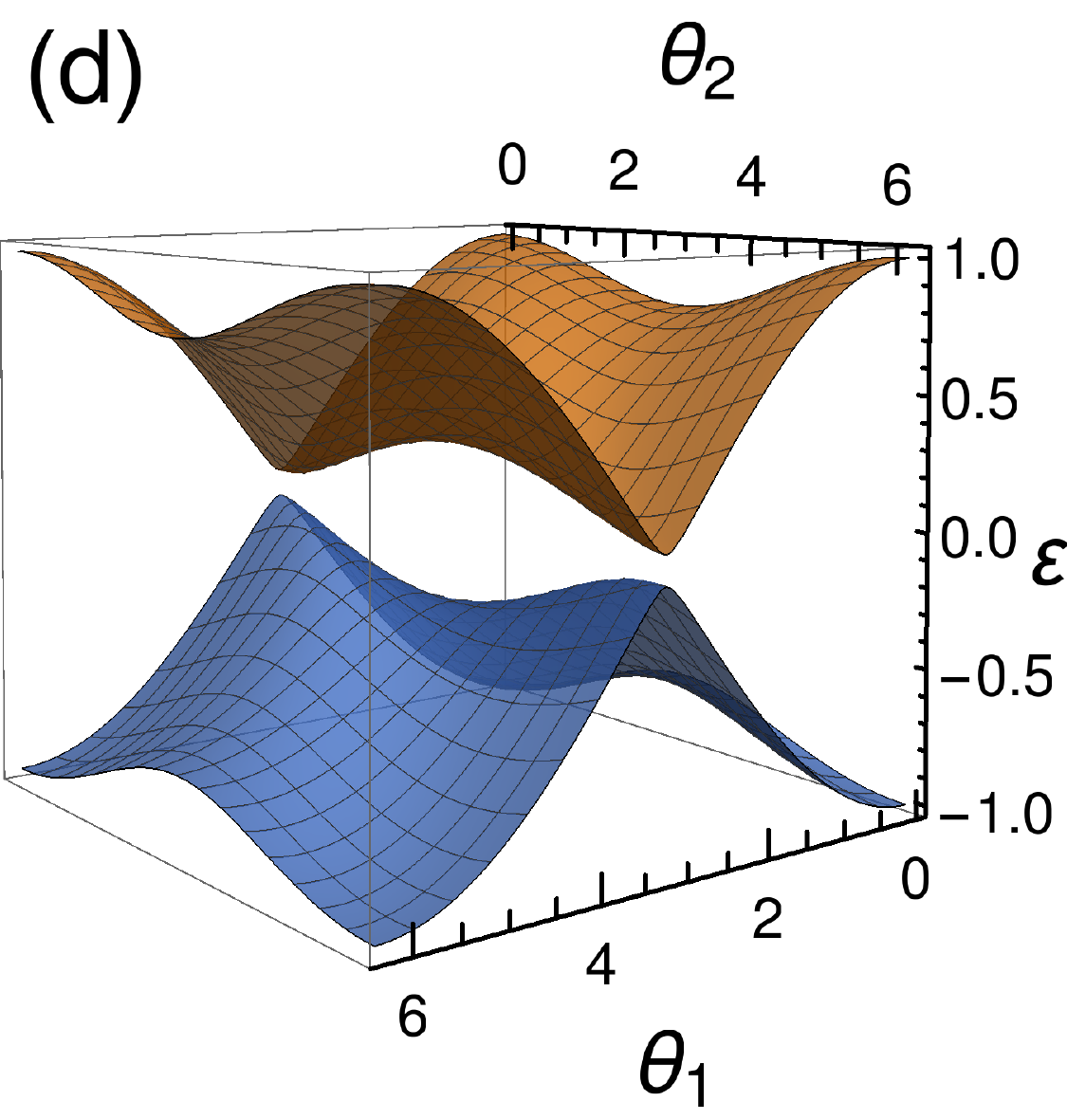} \\
\caption{[Color online] (a) Schematics of a three-terminal Josephson junction. Solid (dash) lines indicate electron (hole) propagation, $\hat{s}(\Phi)$ is the normal-region scattering matrix that can be tuned by external magnetic flux $\Phi$, and $\hat{r}_A$ is the Andreev reflection matrix. (b) Phase diagram for $\varphi=0$. For $b \in [b_{-}(a,0),\, b_{+}(a,0)]$ there is a pair of zero-energy states at $\boldsymbol{\Theta}_{\pm}$, as shown in (c). (c) The example of a gapless ABS spectrum for $a = 0.3$, $b=1/\sqrt{2}$, $\varphi=0$, and (d) gapped spectrum for $a = 0.3$, $b=0.9$, $\varphi=0$.} 
\label{Fig-ABS-trs}
\end{figure}

Most recently, it was proposed that topological properties of various kinds can be effectively engineered and manipulated in multiterminal Josephson junctions (JJs) \cite{Akhmerov,Yokoyama,Riwar,Eriksson,Houzet}. One of the most crucial aspects of this fruitful idea, from the stand-point of its experimental realization, is that such band topology engineering does not require the material constituents forming the junction to be topological. Rather, the topology emerges by design and is harbored by the subgap Andreev bound states (ABS) localized in the junction. The core essence of the idea can be summarized as follows. The ABS spectrum in a two-terminal junction is a periodic function of superconducting phase difference. This is an equivalent to a dispersion relation of a particle in a one-dimensional crystal, where the superconducting phase difference plays the role of momentum and its periodicity modulo $2\pi$ mimics a Brillouin zone. Extending this analogy to a three-terminal junction yields two-dimensional sheets of Andreev levels controlled by two phase differences between superconducting terminals. Remarkably, this system can realize an analog of quantum spin Hall insulator as characterized by a quantized conductance, even though the underlying physics is very different. The four-terminal junctions can further realize three-dimensional Weyl singularities in the ABS spectra that carry topological Berry fluxes. The topological properties of the ABS spectra are determined by the details of the normal-region scattering matrix connecting superconducting leads. However, the precise requirements for the occurrence of band topology are not yet well understood in general. We find some exact analytical results for a particular realization of a scattering matrix from the orthogonal and  unitary symmetries in a limit where each superconducting reservoir is linked by a single conduction channel. We find Weyl singularities in the three-terminal setup when the system lacks time-reversal symmetry and fully explore the topological phase diagram of the model. These results may guide future experimental searches and trigger further theoretical generalizations.   

\textbf{\textit{Scattering matrix formalism}}. Formation of the subgap bound states in the JJs is the result of coherent multiple Andreev reflections that describe electron-to-hole conversion at the superconductor-normal (SN) interface. In transport theory the spectrum of such localized states can be found by the Beenakker's determinant formula \cite{Beenakker-PRL}
\begin{equation}\label{ABS-Det}
\dett\left[1-\gamma(\varepsilon)\hat{r}_A\hat{s}^*(-\varepsilon)\hat{r}^*_A\hat{s}(\varepsilon)\right]=0, 
\end{equation}  
where $\gamma(\varepsilon)=\exp(-2i\arccos\varepsilon)$. For brevity we assume that all superconducting terminals have the same energy gap $\Delta$ and choose to measure energies in units of $\Delta$ so that $\varepsilon$ is dimensionless. We also assume spin-rotation symmetry. Equation~\eqref{ABS-Det} has a transparent physical meaning. Indeed, the diagonal matrix $\hat{r}_A=e^{i\hat{\theta}}$ corresponds to Andreev reflections at the junction interfaces, with $\hat{\theta}=\mathrm{diag}\{\theta_0,\theta_1,\ldots\}$ being the corresponding phases of superconducting terminals, while $\gamma(\varepsilon)$ captures an additional phase shift due to the mismatch of electron and hole quasimomenta. The scattering matrix $\hat{s}(\varepsilon)$ [$\hat{s}^\ast(-\varepsilon)$] describes propagation of electron [hole] -like excitations in the normal region of the junction between superconductors.        

We begin our analysis by a brief recap of essential results that follow from Eq.~\eqref{ABS-Det} in two-terminal junctions. In the RMT limit, which neglects energy dependence of the normal-region scattering matrices, there is one-to-one correspondence between the spin-degenerate energy spectrum of ABS $\varepsilon_k=\pm\sqrt{1-T_k\sin^2\theta/2}$ and transmission eigenvalues of the scattering matrix $T_k$, where the index $k$ labels conduction channels in the junction $k=1,\ldots,N$. For each channel, Andreev levels come in opposite-energy pairs and each level is doubly degenerate as a consequence of the Kramers theorem. The Andreev levels cross at $\theta=\pi$ for perfectly transmitting channels $T_k=1$, while exhibiting avoided crossings for any finite transparency $T_k<1$ with the gap $2\sqrt{1-T_k}$. Physically the RMT limit corresponds to the approximation $L/\xi\to0$, where $L$ is the length of the junction and $\xi$ is the superconducting coherence length, which is justified for point-contact/quantum-dot junctions. Relaxing on this condition leads to the appearance of several qualitatively new features in the spectra of ABS. (i) For a small but finite $L/\xi$, the Andreev levels decouple from the continuum of states at phases $\theta_l=2\pi l$, with $l \in \mathbb{Z}$, which is in contrast to the RMT result $\varepsilon_k(\theta_l)=\pm 1$, and the energy of decoupling $\delta$ is of the order $\delta \sim (L/\xi)^2$. (ii) For a longer junction $L/\xi \gtrsim 1$, the decoupling energy grows and a new pair of levels emerges within the energy window $\varepsilon \in \pm[1-\delta,1]$. (iii) Once $L/\xi \gg 1$ the Andreev levels start to densely populate all the subgap region and form a band with small level spacing. To capture the crossover regime to a long junction one has to employ semiclassical methods based on either the Eilenberger equation for ballistic junctions~\cite{Kopnin} or the Usadel equation for diffusive ones~\cite{AL}. (iv) Inclusion of spin-orbit interaction couples the spin of the bound states to the superconducting phase difference and lifts the Kramers degeneracy of the spectrum. This leads to additional features appearing both at zero energy and at the gap edges. All these complexities attracted much attention recently with a particular  emphasis on three-terminal \cite{Houzet,Akhmerov,Padurariu} and four-terminal \cite{Yokoyama,Riwar,Eriksson,Belzig,Linder,Volkov} junctions.    

\textbf{\textit{Andreev spectra}}. Three-terminal Josephson junctions, as schematically shown in Fig.~\ref{Fig-ABS-trs}(a), are the main focus of our work. Because of the overall gauge invariance, one superconducting terminal can be considered at zero phase $\theta_0=0$ so that the ABS spectrum in the device is controlled by the remaining two phases $\theta_{1,2}$ and particular properties of the scattering matrix $\hat{s}$. Current conservation implies that $\hat{s}$ is a unitary matrix: $\hat{s}^{-1}=\hat{s}^\dag$. Its size is determined by the sum of the numbers of incoming modes in the leads. For simplicity we analyze Eq.~\eqref{ABS-Det} for the energy-independent scattering matrix relevant for the RMT limit. Furthermore, we assume that each superconducting terminal is coupled by only a single conducting channel. 

When the system lacks time-reversal symmetry, unitarity is the only constraint on $\hat{s}$. This corresponds to a circular unitary ensemble in the RMT classification. Thus for a single-channel limit of three-terminal devices under consideration the normal-region scattering matrix $s_{ij}$ has size $3\times3$ and in general can be determined by nine real parameters \cite{Dita}:
\begin{align}\label{s-trsb}
& s_{11} = a e^{i \varphi_{11}}, \, s_{12} = b \sqrt{1-a^2}e^{i \varphi_{12}},  \, s_{31} = \sqrt{(1-a^2)(1-c^2)} e^{i \varphi_{31}}, \nonumber \\ 
& s_{13} = \sqrt{(1-a^2)(1-b^2)} \, e^{i \varphi_{13}}, 
      \quad s_{21} = c \sqrt{1-a^2} \, e^{i \varphi_{21}}, \nonumber \\
& s_{22} = -a b c \, e^{i(\varphi_{12}+\varphi_{21}-\varphi_{11})} + \sqrt{(1-b^2)(1-c^2)} e^{i \varphi_{22}},\nonumber \\ 
&s_{23} = -e^{i \varphi_{13}} \left[ a c \sqrt{1-b^2} \, e^{i(\varphi_{21}-\varphi_{11})}+ b \sqrt{1-c^2} e^{i(\varphi_{22}-\varphi_{12})} \right],  \nonumber \\
& s_{32} = -e^{i \varphi_{31}} \left[ a b \sqrt{1-c^2} \, e^{i(\varphi_{12}-\varphi_{11})}+ c \sqrt{1-b^2} e^{i(\varphi_{22}-\varphi_{21})} \right], \nonumber \\
& s_{33} = e^{i ( \varphi_{13}+\varphi_{31})} \left[ -a \sqrt{(1-b^2)(1-c^2)} \, e^{-i \varphi_{11}} + b c \, e^{i(\varphi_{22} - \varphi_{12}- \varphi_{21})} \right], 
\end{align}
where $a,b,c \in [0,1]$, and $\varphi_{11,22,12,13,21,31} \in [0,2\pi]$. Interestingly, for this case Eq. \eqref{ABS-Det} can be written as a cubic antipalindromic equation $(\gamma-1)(\gamma^2-2B\gamma+1)=0$, which gives a flat-band solution $\varepsilon=1$ and a dispersive band solution
\begin{equation}\label{ABS}
\varepsilon(\theta_1,\theta_2) = \sqrt{\frac{B(\theta_1,\theta_2)+1}{2}},
\end{equation} 
where the $B$ function reads
\begin{eqnarray} \label{B-trsb}
&&B= \frac{1}{2}\left[ 2 a^2 - (1+a^2)(b^2+c^2 - 2b^2c^2)\right. \nonumber\\ 
&&-\left. 4 abc \sqrt{(1-b^2)(1-c^2)} \cos\varphi \right] \nonumber\\ 
&&+ bc(1-a^2) \cos\vartheta_1 + (1-a^2) \sqrt{(1-b^2)(1-c^2)} \cos\vartheta_2  \nonumber\\
&&+\left[bc (1+a^2) \sqrt{(1-b^2)(1-c^2)} + a (b^2+c^2 - 2b^2c^2) \cos\varphi \right]\nonumber \\ 
&&\times \cos(\vartheta_1-\vartheta_2)+a (b^2-c^2) \sin\varphi \sin(\vartheta_1-\vartheta_2).  
\end{eqnarray}
Here $\vartheta_{1,2}=\theta_{1,2}+\phi_{1,2}$ are shifted superconductor phases
with $\phi_{1} \equiv \varphi_{12} - \varphi_{21}$, $\phi_{2} \equiv \varphi_{13} - \varphi_{31}$,  and 
$\varphi \equiv \varphi_{11} + \varphi_{22} - \varphi_{12} -\varphi_{21}$.
Consequently, there are only six independent parameters of the scattering matrix $\{ a, b, c, \varphi, \phi_{1,2} \}$ that enter the spectrum of ABS. Furthermore, $\phi_{1,2}$ only shift the phases of the leads. Each band has its mirror image at $\varepsilon\to-\varepsilon$.

The presence of time-reversal symmetry imposes additional constraints. In particular if in addition spin-rotation symmetry is present, which corresponds to a RMT circular orthogonal ensemble, then the scattering matrix is unitary and symmetric: $\hat{s}=\hat{s}^T$. For a $3\times3$ matrix this implies only six independent real parameters, which can be reduced from the parametrization in Eq. (\ref{s-trsb}) by setting $c=b$ and $\phi_{1,2}=0$, so that the $B$ function in Eq. \eqref{B-trsb} is simplified to 
\begin{eqnarray}\label{B-trs}
&&B=a^2+ (1-a^2) \left[ b^2 \cos{\theta_1}  + (1-b^2) \cos{\theta_2} \right]\nonumber\\  
&&- 2  b^2 (1-b^2) (1+ a^2 + 2 a \cos{\varphi})\sin^2 \left( \frac{\theta_1-\theta_2}{2} \right),
\end{eqnarray}   
while the ABS spectrum is still given by Eq. \eqref{ABS}. This particular limit admits a complete analytical solution. The Andreev energy spectrum has six potential stationary points: $\bm{\Theta}_1 = (0,0), \bm{\Theta}_2 = (\pi,\pi), \bm{\Theta}_3 = (\pi,0), \bm{\Theta}_4 = (0, \pi), \bm{\Theta}_{+} = (\theta_1^0, \theta_2^0), \quad  \bm{\Theta}_{-} = (2\pi-\theta_1^0, 2\pi- \theta_2^0),$ where
$\theta_1^0 = \arccos\left[ \frac{2 a^2 - F_1(a,b,\varphi)}{b^2 \, F_2(a,b,\varphi)} \right]$ and 
$\theta_2^0 = 2\pi - \arccos\left[ \frac{-1 - a^4 + F_1(a,b,\varphi)}{\left( 1-b^2 \right) \, F_2(a,b,\varphi)}\right]$,
with $F_1(a,b,\varphi) = -2 a (1-2b^2) \cos{\varphi} (1  + a^2 + a\cos{\varphi}) + (1+a^2)^2 b^2$, and $
F_2(a,b,\varphi) = (1-a^2)(1 + 2a\cos{\varphi} + a^2)$ so that $\theta_1^0 \in [0, \pi]$ and $\theta_2^0 \in [\pi, 2\pi]$. In general $\bm{\Theta}_1$ is the maximum point with energy $\varepsilon( \bm{\Theta}_1) = 1$ and $\bm{\Theta}_2$ a saddle point with $\varepsilon( \bm{\Theta}_2) = a$. For convenience we introduce the functions 
\begin{eqnarray}
b_{+}(a, \varphi) = \sqrt{\frac{1 + a \cos{\varphi} }{1 + 2 a\cos{\varphi} + a^2 }}, \quad b_{-}(a, \varphi)=\sqrt{a}b_+ (a, \varphi),
\end{eqnarray}
such that $b_+ (a, \varphi) \in [1/\sqrt{2}, \, 1]$ and $b_+ (a, \varphi) \in [0,\,1/\sqrt{2}]$ as $a$ changes in a range $a\in[0,1]$ for a fixed value of $\varphi$. When $b \in [b_+ (a, \varphi), \, 1]$, $\bm{\Theta}_3$ is the minimum point with energy 
\begin{equation}
\varepsilon(\bm{\Theta}_3) = \sqrt{1-2 b^2 + \left(1+ a^2\right) b^4 - 2 a  b^2 \left(1-b^2\right) \cos{\varphi}},
\end{equation}
$\bm{\Theta}_4$ is a saddle point, and no solution exists for $\bm{\Theta}_\pm$. When $b \in [0, \, b_- (a, \varphi)]$, $\bm{\Theta}_4$ is the minimum point with energy
\begin{equation} 
\varepsilon(\bm{\Theta}_4) = \sqrt{b^4-\left(1-b^2\right) \left[ a^2\left(1-b^2\right) + 2 a  b^2 \cos{\varphi} \right] },
\end{equation}  
$\bm{\Theta}_3$ is a saddle point, and no solution exists for $\bm{\Theta}_\pm$. Finally, when 
$b \in [b_- (a, \varphi), \, b_+ (a, \varphi)]$, $\bm{\Theta}_\pm$ are such that
\begin{equation} \label{TRS-mass}
\varepsilon(\bm{\Theta}_\pm) = \frac{a \left| \sin{\varphi} \right|}{\sqrt{a^2+2 a \cos{\varphi}+1}},
\end{equation}
and $\bm{\Theta}_{3,4}$ are saddle points. When $\varphi=0$ there is a pair of zero-energy states at $\bm{\Theta}_\pm$, as shown in Fig.~\ref{Fig-ABS-trs}(c). As $\varphi$ passes through zero, the energy gap closes and reopens. The gapped phase is shown in Fig.~\ref{Fig-ABS-trs}(d), whereas the phase diagram in a parameter space of $\{a,b\}$ is shown in Fig. \ref{Fig-ABS-trs}(b). 

We can further derive an effective low-energy Hamiltonian by expanding the Andreev spectrum about $\bm{\Theta}_\tau$ ($\tau = \pm$), which takes the form of massive Dirac fermions in two dimensions $\hat{H}_\tau = \hat{\bm{V}_\tau} \cdot \bm{P} + M \hat{\sigma}_3$,
where effective momentum $\bm{P}$ is a rotation of $\delta\bm{\Theta}\equiv \bm{\Theta} - \bm{\Theta}_\tau$: $\bm{P}=\hat{R}_\tau(a,b) \delta \bm{\Theta}$, $\hat{\sigma}_{i}$ are Pauli matrices operational in the basis of the two degenerate states, and $\hat{\bm{V}}_\tau=(v_{\tau,1}(a,b) \hat{\sigma}_1,v_{\tau,2}(a,b) \hat{\sigma}_2)$ the effective velocity. The rotation matrix $\hat{R}_\tau$ and velocity components are determined by the eigenproblem of the matrix $\hat{C}_\tau$ with $C_{\tau, ij} \equiv \frac{1}{4}\partial_{\theta_i} \partial_{\theta_j} B(\varphi=0)|_{\bm{\Theta}=\bm{\Theta}_\tau}$: $\hat{R}_\tau \hat{C}_\tau \hat{R}_\tau^{-1} = \mathrm{diag}\{v_{\tau,1}^2, v_{\tau,2}^2 \}$. Finally, the Dirac mass $M(a,b,\varphi) = \varepsilon(\bm{\Theta}_\pm)$ [Eq.~(\ref{TRS-mass})] is positively defined for any phase value of the scattering matrix so that this case is topologically trivial.

\begin{figure}
\includegraphics[width=0.23\textwidth]{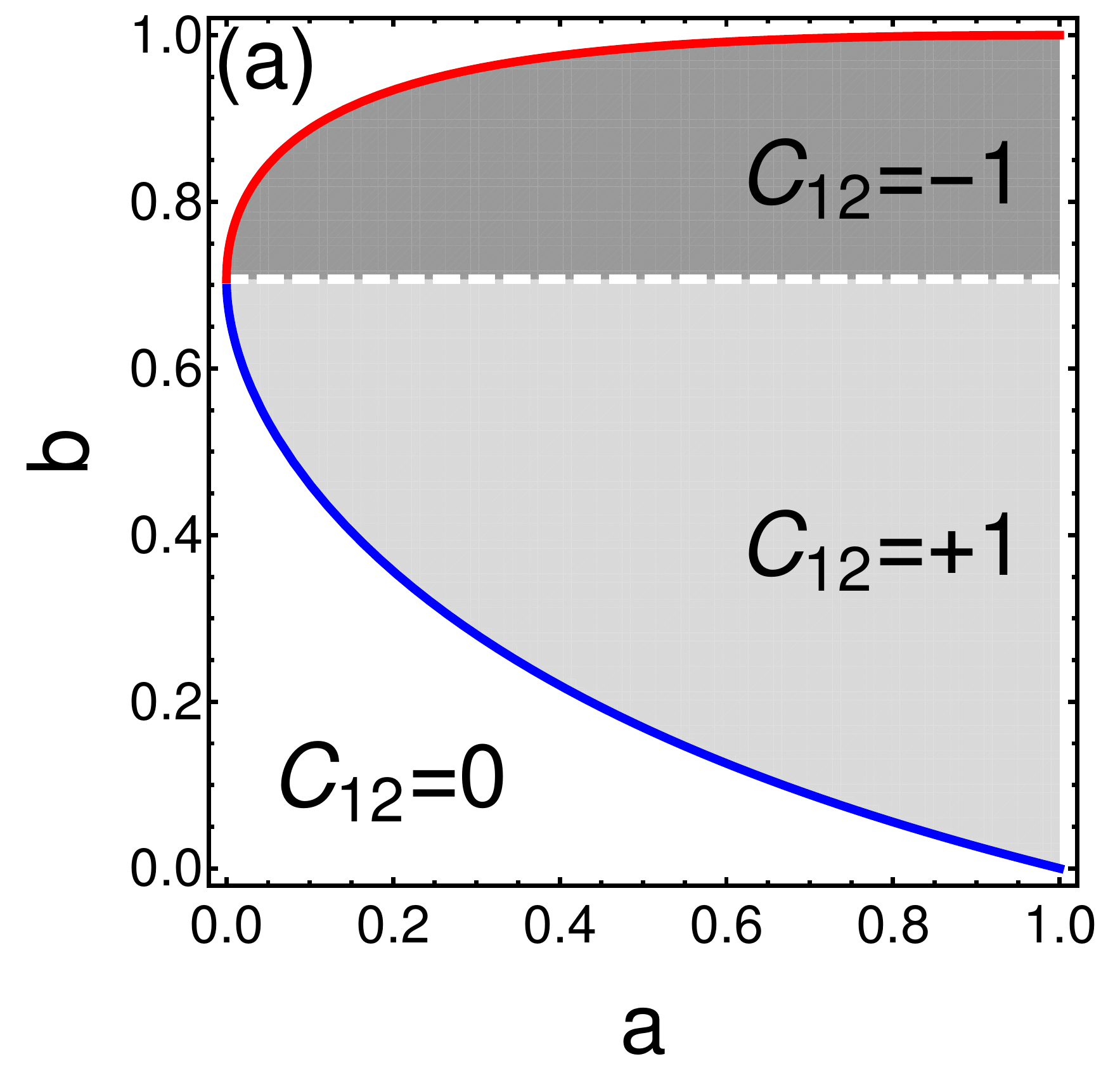}
\includegraphics[width=0.23\textwidth]{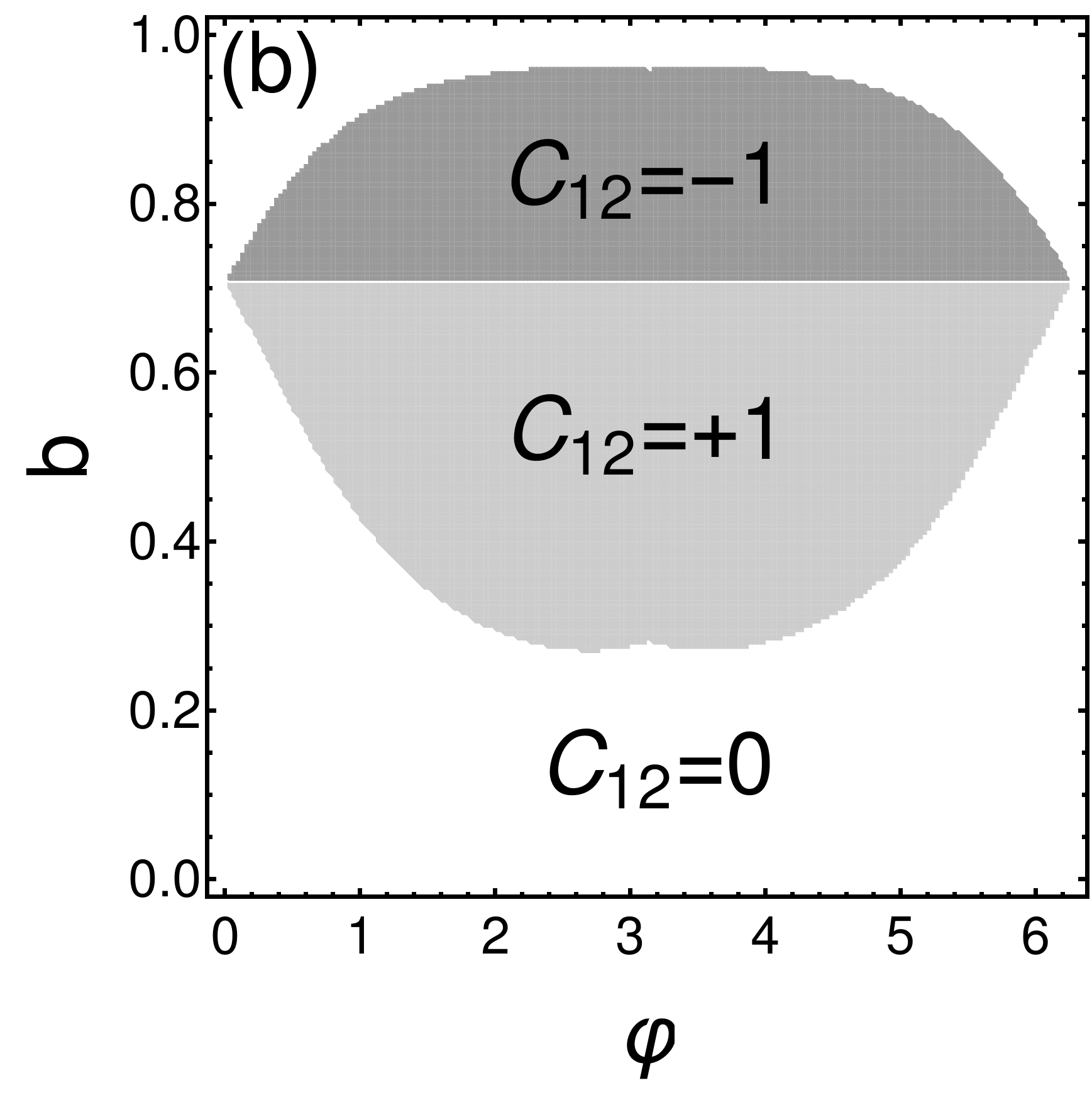}
\caption{Phase diagram and the Chern number $C_{12}$ of Andreev bounds states for three-terminal Josephson junctions in the case of broken time-reversal symmetry.  We take $b = \sqrt{1-c^2}$ for simplicity.  $C_{12}$ as a function of parameters of the scattering matrix: (a) $a$ and $b$ for $\varphi = \pi$, and (b) $b$ and $\varphi$ for $a=0.3$, respectively.} 
\label{Fig-chern}
\end{figure}

\begin{figure}
\includegraphics[width=0.23\textwidth]{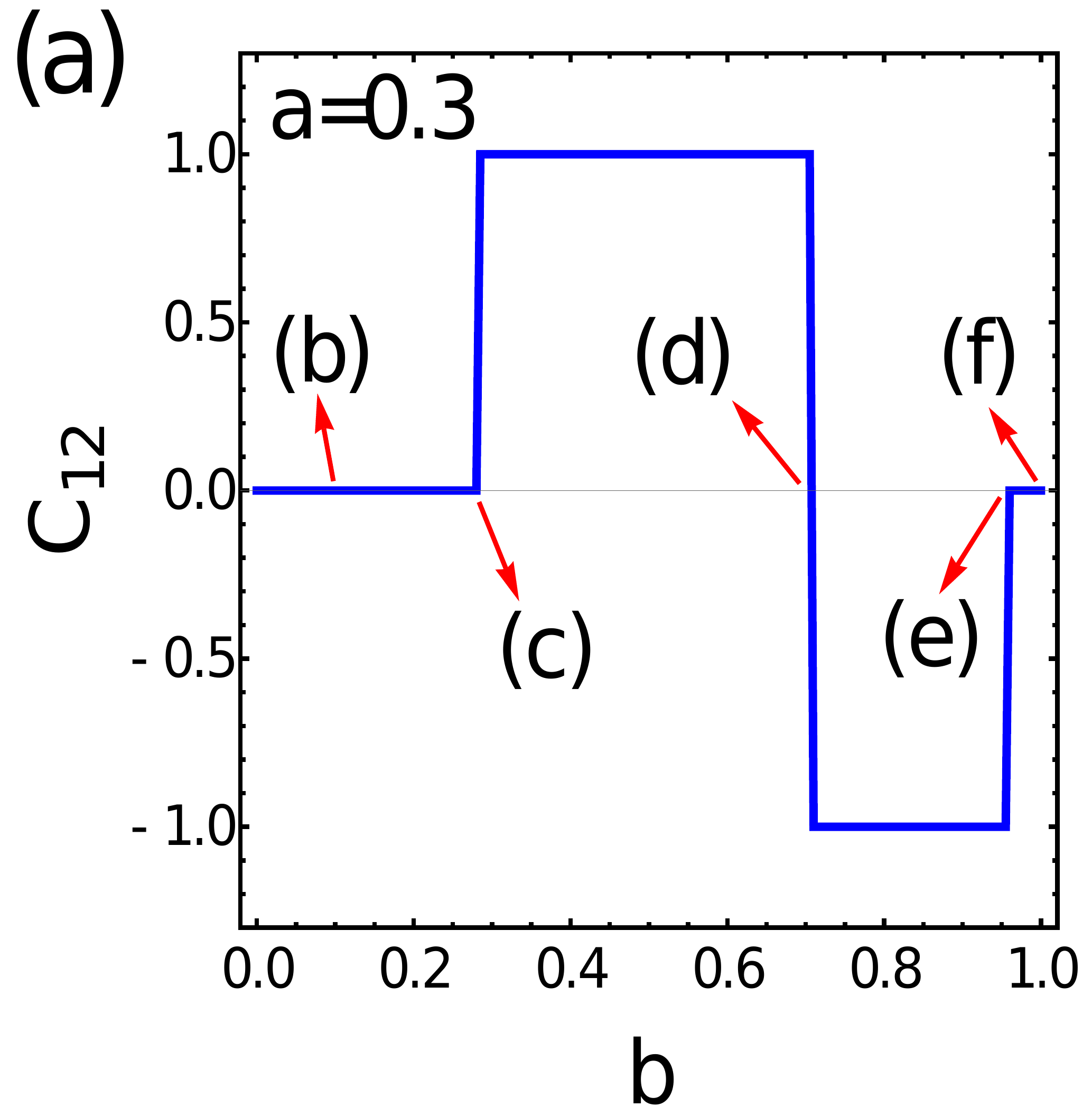}
\includegraphics[width=0.23\textwidth]{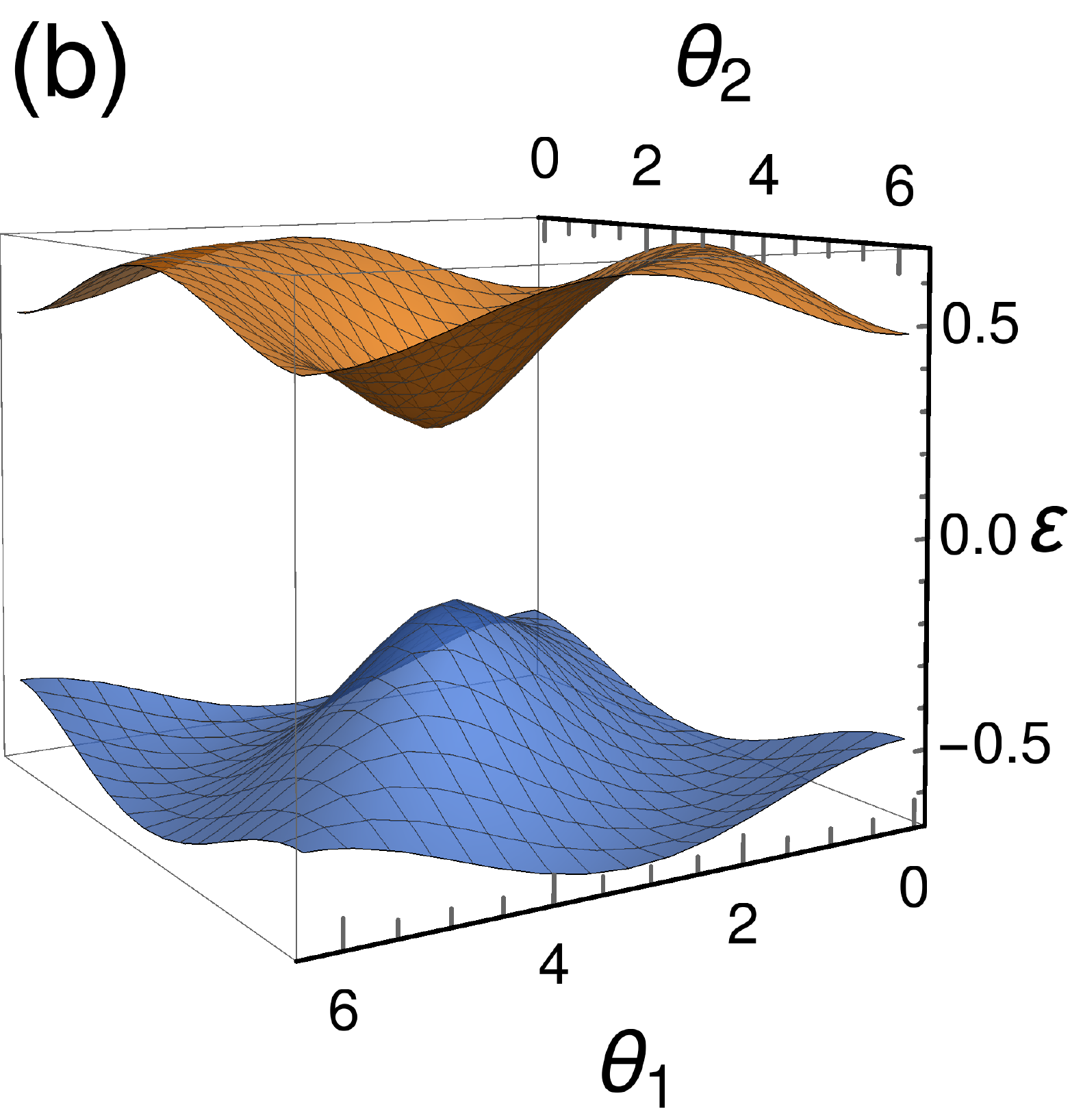} \\
\includegraphics[width=0.23\textwidth]{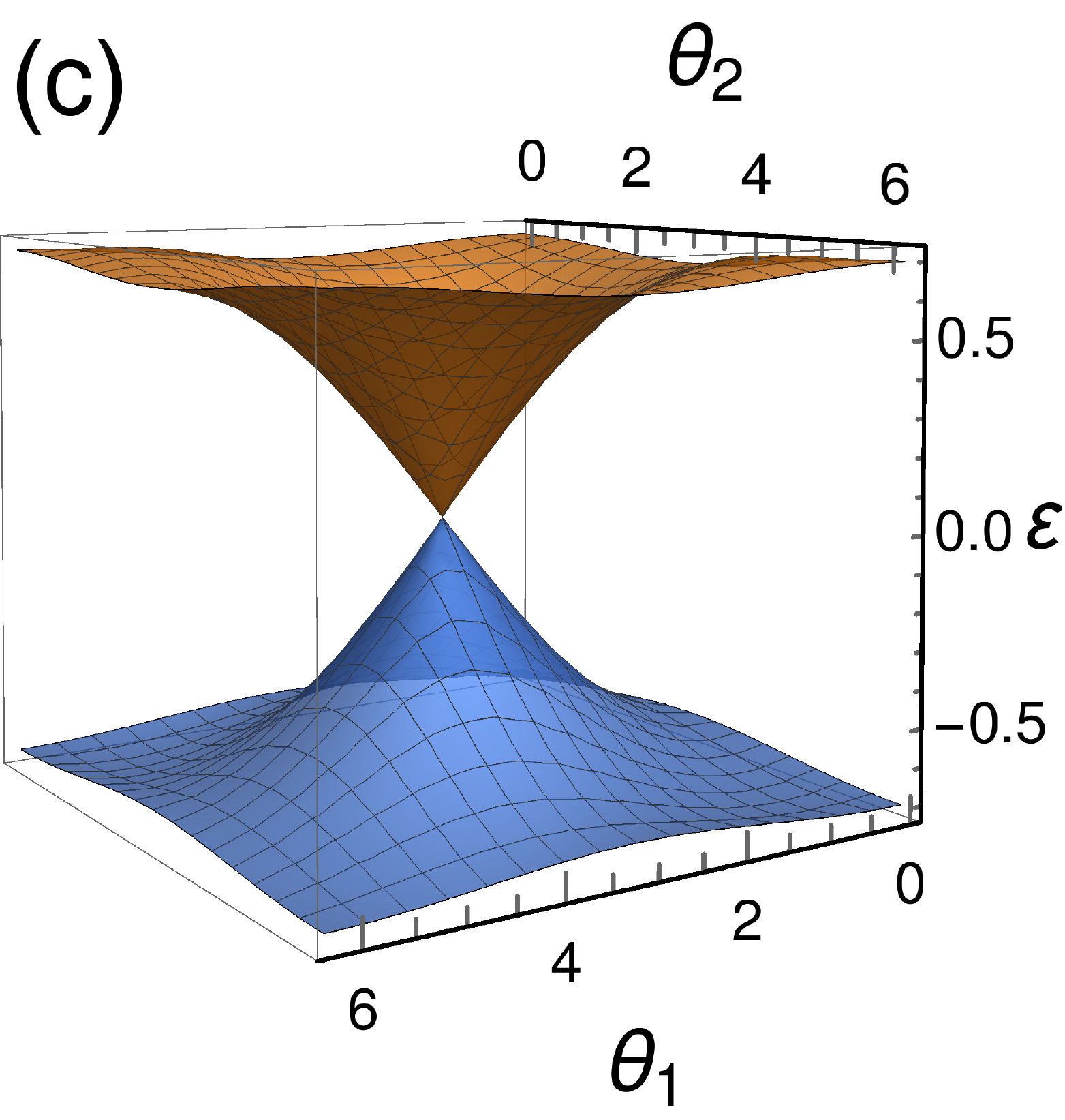} 
\includegraphics[width=0.23\textwidth]{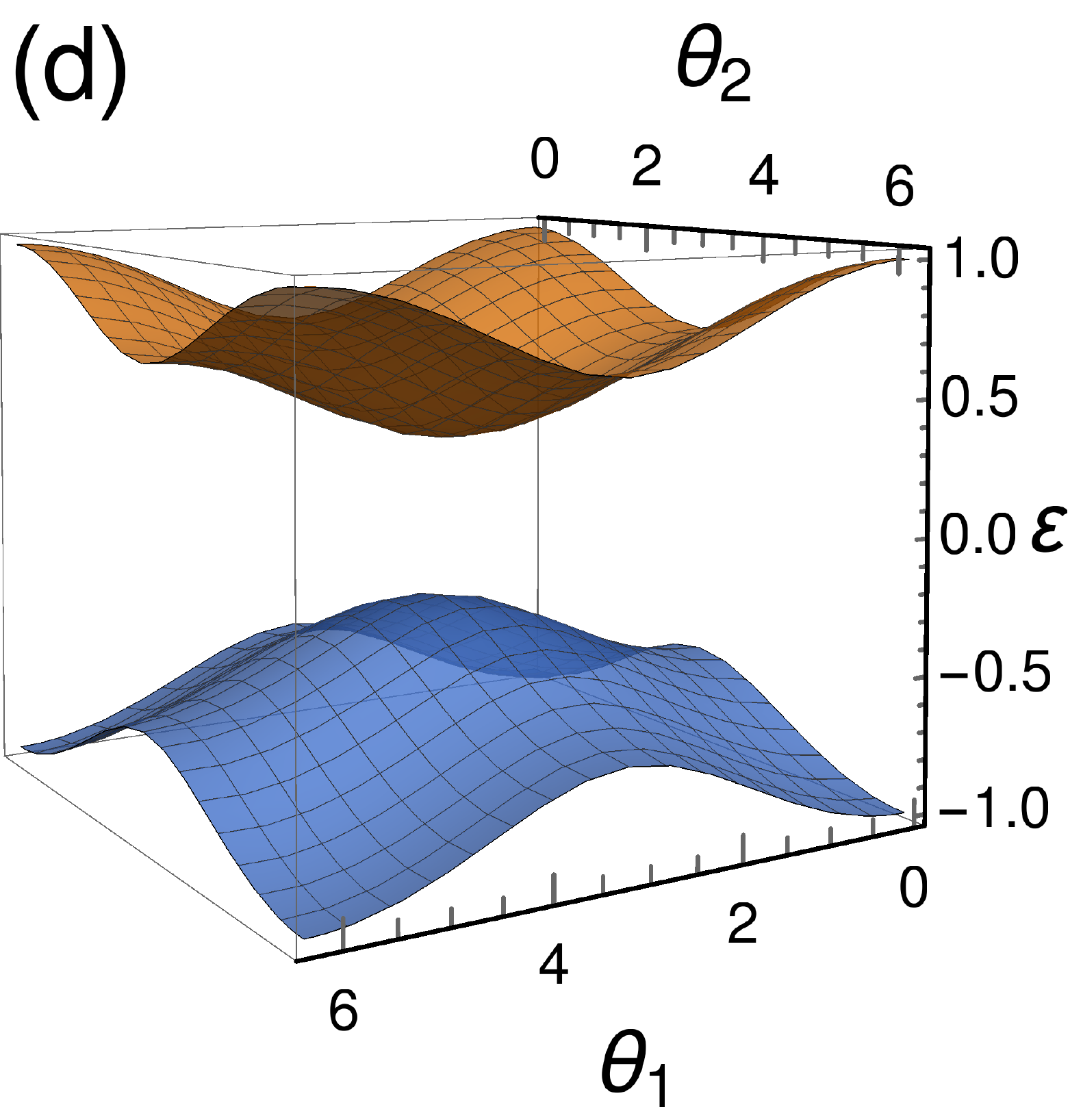}  \\
\includegraphics[width=0.23\textwidth]{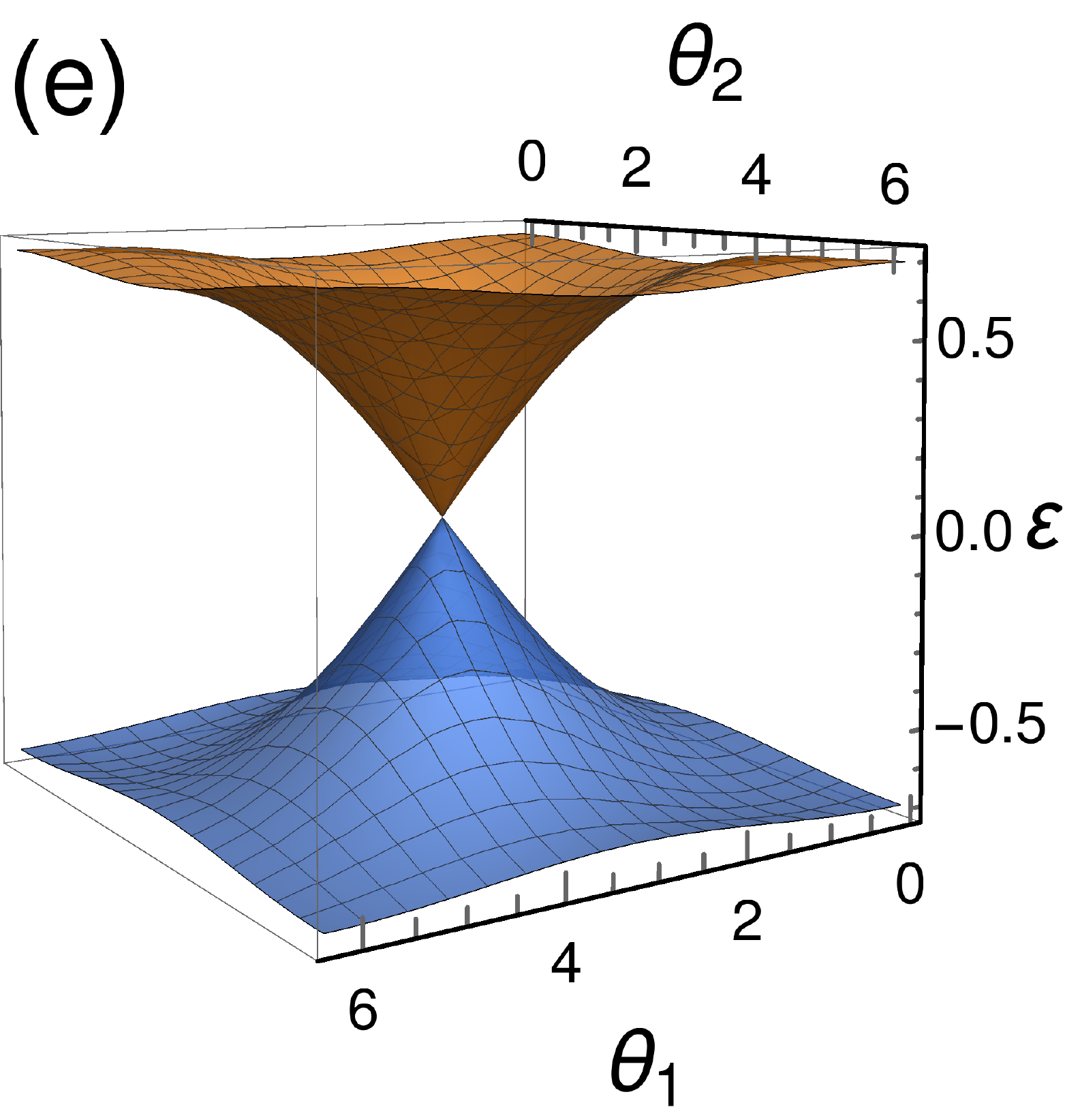}
\includegraphics[width=0.23\textwidth]{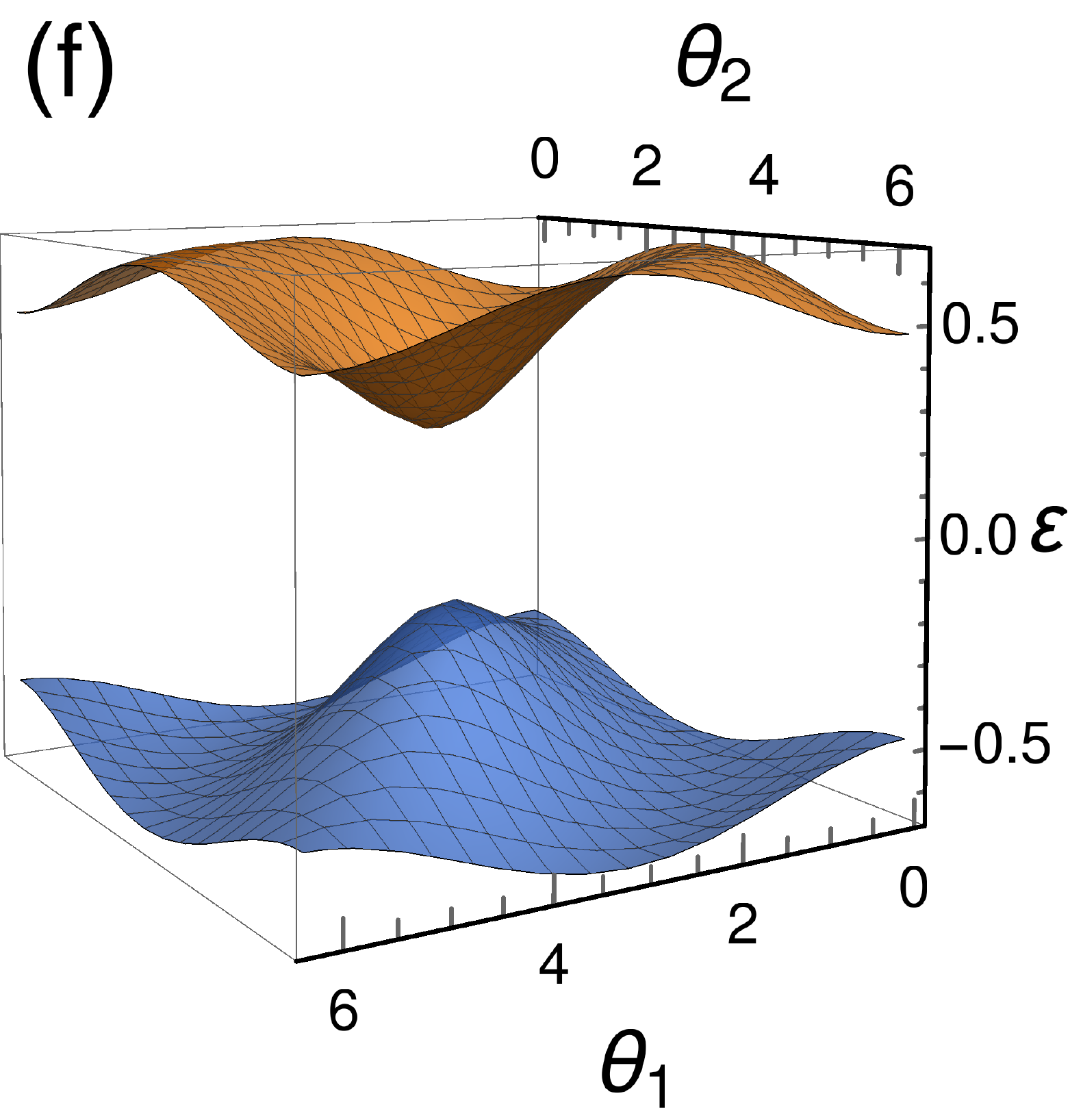}
\caption{Energy spectrum of Andreev bounds states for three-terminal Josephson junctions in the case of unitary symmetry of the scattering matrix in the normal region of the junction. We take the same parameters as those in Fig.~\ref{Fig-chern}(a) and fix $a=0.3$. (a) Chern number as a function of $b$, and (b)-(f) the Andreev spectra at $b=0.1$, $b_-$, $b_0$, $b_{+}$, and $0.99$, indicted in (a).} 
\label{Fig-spectra}
\end{figure}

In contrast, in the absence of time-reversal symmetry the ABS bands become topologically nontrivial, in particular, due to the condition $b \neq c$. This most interesting scenario can be realized by applying a magnetic flux piercing the normal junction area~\cite{Houzet}. Depending on the choice of parameters and fluxes in our model we find very rich behavior of the energy bands. For a special case $b = \sqrt{1-c^2}$ and $\varphi=\pi$ the spectrum can be studied analytically and reveals nontrivial topology, as exemplified in Figs.~\ref{Fig-chern} and \ref{Fig-spectra}. Weyl points appear at $\vartheta_1^0 = \vartheta_2^0 = \pi$ for $b= b_\pm(a)$ with $b_\pm(a) \equiv \frac{1 \pm \sqrt{a}}{2 \sqrt{1+a}}$. We define $b_0=1/\sqrt{2}$ representing the time-reversal-symmetric point. The Chern number for the bands of ABS can be computed according to the standard procedure by integrating Berry curvature over the unit cell spanned by phases $\theta_{1,2}$~\cite{Riwar,Eriksson,Houzet},
\begin{equation}
C_{12} = \frac{1}{ 2 \pi} \iint_{0}^{2\pi} \! d\theta_{1} d\theta_{2} \, B_{12}, \quad B_{12} =  -2 \sum_{k} \mathrm{Im} \langle  \partial_{\theta_{1}} \psi_k | \partial_{\theta_{2}} \psi_k \rangle, 
\end{equation} 
where $B_{12}$ is the Berry curvature with $|\psi_k  \rangle$ being the bound state $k$. For our model the Chern number as a function of $a$ and $b$ reads [see Fig. \ref{Fig-chern}(a)]
\begin{equation}\label{C12}
C_{12} =  \begin{cases} &\,   \phantom{+}0, \quad   a \in [0, b_-) \cup (b_+,1], \\
                        &\,  +1, \quad   a \in ( b_-, b_0), \\
                         &\, -1, \quad   a \in ( b_0, b_+). \\
\end{cases}
\end{equation}
We note that the Chern number vanishes $C_{12}=0$ at the time-reversal-symmetric point $b=b_0$ and takes opposite signs $C_{12} = \mathrm{sgn}(b_0-b)$ for $b \sim b_0$. In Fig.~\ref{Fig-chern}(b) we also show the numerical result for the phase diagram of trivial-to-topological quantum phase transitions as a function of $b$ and $\varphi$ for a fixed parameter $a$. 

\textbf{\textit{Conductance and Chern numbers}}. As shown in Ref. \cite{Riwar}, the existence of Weyl points in the multiterminal JJs can be probed by nonlocal conductance measurements that are expected to be quantized in the topological regime. Indeed, the current flowing into the first lead as a result of applied subgap voltage $eV_2\ll\Delta$ to the second lead is of the form 
\begin{equation}\label{I}
I_1(t)=\frac{2e \Delta}{\hbar}\partial_{\theta_1}\varepsilon-2e\dot{\theta}_2B_{12}
\end{equation}   
where by virtue of the second Josephson equation $\dot{\theta}_2=2eV_2/\hbar$. The first term in Eq. \eqref{I} corresponds to the adiabatic current, and the second term is the first-order correction that is in a way an anomalous velocity component governed by the Berry curvature $B_{12}$. In this sense, the instantaneous current can be used to directly assess the Berry curvature \cite{Avron,Gritsev,Xu,Martinis,Lehnert}. When two incommensurate voltages are applied to both leads, the two phases uniformly sweep an effective Brillouin zone of the ABS band structure. In the dc limit the adiabatic current averages out to zero, whereas the anomalous velocity component is replaced by its average value. As a result, the current is linear in the voltages $\bar{I}_\alpha=G_{\alpha\beta}V_\beta$ and conductance is defined by the Chern number 
\begin{equation}
G_{12}=-\frac{4e^2}{h}C_{12}, 
\end{equation}        
with $C_{12}$ taken from Eq. \eqref{C12} within our model. A particular example is depicted in Fig. \ref{Fig-spectra}(a). The corresponding shapes of Andreev bands are displayed in Fig. \ref{Fig-spectra}(b-f). 
We do not delve into a detailed discussion of the conditions required for observability of quantized conductances as both Landau-Zener nonadiabatic conditions and inelastic relaxation processes play an important role. This analysis was carried out in Ref. \cite{Eriksson} for the four-terminal setup with an estimate that topological quantization becomes visible for voltages of the order $\lesssim 10^{-2}\Delta/e$.   

\textbf{\textit{Summary and outlook}}. We considered a simple model of a three-terminal Josephson junction that realizes the band topology of subgap Andreev levels. Weyl singularities appear in the spectrum when the system lacks time-reversal symmetry. The latter is captured by the properties of the scattering matrix of the normal region connecting superconducting leads and can be tuned by external magnetic flux piercing the junction area. The topological regime is quantified by nonvanishing Chern numbers that translate into a quantized nonlocal conductance \cite{Note}. Three- and four-terminal Josephson junctions have been recently realized in experiments \cite{Kouwenhoven,Giazotto-1,Giazotto-2}. These advances open new avenues not only to study new physics of topological mesoscopic superconducting systems, but also to explore opportunities in implementing these multiterminal devices into superconducting qubits to seek topological protection in quantum computation, high fidelity gates, and potentially braiding operations by voltage pulses. It is also important to clarify how such artificial multiterminal ``materials'' fit into the standard periodic table of topological semimetals as they are conceptually distinct. In terms of transport theories it is of interest to investigate whether multiterminal JJs may also provide an alternative platform to study properties of Weyl semimetals related to chiral anomaly both within and beyond the linear response.  

\textbf{\textit{Acknowledgment}}. We are grateful to Vladimir Manucharyan and Yuli Nazarov for fruitful discussions. This work was financially supported by NSF Grant No. DMR-1606517 and in part BSF Grant No. 2014107 (H.X.), by NSF CAREER Grant No. DMR-1653661, NSF EAGER Grant No. DMR-1743986, and Vilas Life Cycle Professorship program (A.L.), by ARO Grant No. W911NF-15-1-0248 (M.V.), and the Wisconsin Alumni Research Foundation.
  

\end{document}